\documentclass[aps,prl,twocolumn,showpacs,superscriptaddress,groupedaddress]{revtex4}  
\usepackage{graphicx}  
\usepackage{dcolumn}   
\usepackage{bm}        
\usepackage{amssymb}   
\usepackage{amsmath}
\usepackage{epstopdf}

\def \Nt {{N_{\tau}}}
\def \Nc {{N_{c}}}

\newcommand{\expval}[1]{\left\langle #1 \right\rangle}

\newcommand{\pbp}{\bar{\psi}\psi}
\newcommand{\p}{\psi}
\newcommand{\pb}{\bar{\psi}}
\newcommand{\dpdpb}{d\psi d \bar{\psi}}

\newcommand{\lr}[1]{\left( #1 \right)}

\def \tr {{\rm tr}}

\def \bet0{\beta_0}
\def \bet1{\beta_1}
\def \simgt{\,\rlap{\lower 7.5 pt\hbox{$\mathchar \sim$}}\raise 3 pt \hbox{$>$}\,}
\def \simlt{\,\rlap{\lower 7.5 pt\hbox{$\mathchar \sim$}}\raise 3 pt \hbox{$<$}\,}

\def\lsim{\raise0.3ex\hbox{$<$\kern-0.75em\raise-1.1ex\hbox{$\sim$}}}
\def\gsim{\raise0.3ex\hbox{$>$\kern-0.75em\raise-1.1ex\hbox{$\sim$}}}

\newcommand{\hnu}{{\hat{\nu}}}

\newcommand{\Ord}[1]{\mathcal{O}\left( #1 \right)}

\begin{document}


\title{The lattice QCD phase diagram in and away from the strong coupling limit}

\date{\today}

\author{Ph.~de Forcrand}

\affiliation{CERN, Physics Department, TH Unit, CH-1211 Geneva 23, Switzerland}

\affiliation{Institut f\"ur Theoretische Physik, ETH Z\"urich, CH-8093 Z\"urich, Switzerland}

\author{J.~Langelage}

\affiliation{Institut f\"ur Theoretische Physik, ETH Z\"urich, CH-8093 Z\"urich, Switzerland}

\author{O.~Philipsen}

\author{W.~Unger}

\affiliation{Institut f\"ur Theoretische Physik, Goethe-Universit\"at Frankfurt, 60438 Frankfurt am Main, Germany}

\begin{abstract}
We study 
lattice QCD with four flavors of staggered quarks.
In the limit of infinite gauge coupling, ``dual'' variables can be
introduced, which render the finite-density sign problem mild and allow
a full determination of the $\mu-T$ phase diagram by Monte Carlo simulations,
also in the chiral limit. However, the continuum limit coincides with the
weak coupling limit. We propose a strong-coupling expansion approach towards
the continuum limit. We show first results, including the phase diagram and
its chiral critical point, from this expansion truncated to next-to-leading order.
\end{abstract}


\maketitle



The properties of QCD as a function of temperature $T$ and matter density are summarized by its phase diagram, whose determination is a major goal of large-scale heavy-ion experiments. Although the
quark-gluon plasma has been observed at high temperature, further features of the phase diagram,
especially a possible QCD critical point, have not been identified yet. On the theory side,
heroic efforts have been devoted to numerical lattice simulations, which are the appropriate tool
for  non-perturbative phenomena like phase transitions. 
However, the fermion determinant becomes complex upon turning on a chemical potential $\mu$
coupled to the quark (or baryon) number. This so-called ``sign problem'' requires 
prohibitively
large computer resources growing exponentially with the lattice 4-volume. Approaches to circumvent this
problem are applicable when $\mu/T \lesssim 1$ only~\cite{Forcrand2009}, and
results on the QCD critical
point are inconclusive. 
%
We want to make progress on this problem by means of 
a strong coupling expansion, as applied to zero density in the early days of lattice gauge theory 
or recently to finite temperature and density with heavy quarks \cite{Fromm2011b,Fromm2012}.
Here we want to address the opposite, chiral limit with a different strategy \cite{Rossi1984,Wolff1985}.
Note that both for heavy and chiral quarks, the strong coupling approach gives access also
to the cold and dense regime of nuclear matter \cite{Fromm2010,Fromm2012}.

The sign problem occurs when elements 
$\langle \psi_i | \exp(-\delta\tau H) | \psi_j\rangle$ of the transfer matrix between 
states $|\psi_i\rangle$ and $|\psi_j\rangle$ sampled by Monte Carlo become negative.
This problem is representation-dependent: if we could work in an eigenbasis of the Hamiltonian,
all matrix elements would be non-negative. Thus, the sign problem will become milder if
we can express the partition function in terms of approximate eigenstates. Now, we know
that QCD eigenstates are color singlets. Therefore, 
instead of performing Monte Carlo on colored gauge links, as done in the usual
approach where fermion fields are integrated out, we integrate
the gauge links {\em first}, and work with the resulting color singlets. This strategy
becomes particularly practical in the strong coupling limit, as we explain below.
In this regime, we reexpress the partition function as a sum over configurations of hadron 
worldlines, similar to the ``dual variables'' used in~\cite{Mercado2013}. 
The resulting sign problem is extremely mild, which allows us to simulate
large lattices at arbitrarily large chemical potentials, and reliably obtain the
full QCD phase diagram. Of course, in the strong coupling limit 
$g\to\infty, \beta=2N_c/g^2 \to 0$ (for $\Nc$ colors),
the lattice is maximally coarse, whereas the continuum limit coincides with the weak coupling limit $g\to 0, \beta\to\infty$. 
In this letter,
we first summarize and clarify the $\beta=0$ phase diagram and then explain
how to include the first, ${\cal O}(\beta)$ corrections, which allows us to
measure Wilson loops at $\beta=0$ and fermionic observables at ${\cal O}(\beta)$.
We then present the QCD phase diagram for small $\beta>0$. 
For $\mu=0$ where we can crosscheck with the full Monte Carlo approach, perfect agreement
is found for small $\beta$.

We adopt the staggered fermion discretization and the Wilson plaquette action with the
partition function:
\begin{eqnarray}
Z_{\rm QCD} &\!\!=\!&\! \!\int \!\!\dpdpb dUe^{S_G+S_F}\!,\;
S_G \!=\! \frac{\beta}{2N_c} \! \sum_P \tr[U_P\!+\!U_P^\dagger] \label{QCDPF}
\\
S_F &\!\!=\!& am_q\sum_x\pb_x \p_x + \frac{1}{2}\sum_{x,\nu} \eta_\nu(x) \gamma^{\delta_{\nu 0}} \label{QCDAction}\\
&\!\!\times\!& 
\!\left[ \pb_x e^{a_t \mu \delta_{\nu 0}} U_\nu(x)\p_{x\!+\!\hat{\nu}} - \pb_{x\!+\!\hat{\nu}} e^{-a_t \mu \delta_{\nu 0}}U_\nu^\dagger(x)\p_x \right]\nonumber
\end{eqnarray}

\noindent
with $a$ and $a_t$ the spatial and temporal lattice spacings, $\gamma$ the anisotropy by which one may tune
$a/a_t$, $m_q$ the quark mass and $\mu$ the quark chemical potential 
(the baryon chemical potential is $\mu_B\!=\!N_c\mu$).  
The $\eta$'s are the usual $\pm 1$ staggered phases.
In the continuum limit $g\rightarrow 0$, our action describes QCD with 4 mass-degenerate quark species.
In the opposite, strong coupling limit $g\rightarrow \infty$, the plaquette, 4-link coupling $\beta$ vanishes
and so does the gauge action $S_G$. Then, the integration over the links 
$U_\nu(x)$ factorizes into a product of one-link integrals which can be 
carried out analytically \cite{Eriksson1981}.
Finally, one performs the Grassmann integration over the fermion fields 
$\p(x), \pb(x)$, and obtains the partition function 
in terms of color-singlet, hadronic degrees of freedom (mesons and baryons) \cite{Rossi1984},
as a sum over discrete graphs on the lattice:
\begin{eqnarray}
Z_{SC} &=& \sum_{\{n,k,\ell\}} \prod_{x} w_x \prod_b w_b \prod_\ell w_\ell \label{eq:Z_SC}  \\
w_x &=& \frac{N_c!}{n_x!}(2am_q)^{n_x}; \quad w_b = \frac{(N_c-k_ b)!}{N_c!k_b!} \label{eq:w_SC}\;.
\end{eqnarray}
The mesons are represented by monomers $n_x\in\{0,\ldots N_c\}$ on sites $x$ 
and dimers $k_b\in\{0,\ldots N_c\}$ on bonds $b=(x,\hnu)$, whereas
the baryons are represented by oriented self-avoiding loops $\ell$.
The weight $w_\ell$ of a baryonic loop $\ell$ and its sign depend on the loop geometry
\cite{Karsch1989}.
Configurations $\{n,k,\ell\}$ must satisfy at each site $x$ the constraint
inherited from Grassmann integration:
\begin{equation}
n_x+\sum_{\hnu=\pm\hat{0},\ldots, \pm \hat{d}}\lr{k_{\hnu}(x) + \frac{N_c}{2} |\ell_\hnu(x)|} = N_c.
\label{Grassmann}
\end{equation}
\noindent
Due to this constraint, mesonic degrees of freedom (monomers and dimers) cannot occupy baryonic sites.

\begin{figure}[t!]
\includegraphics[width=0.49\textwidth]{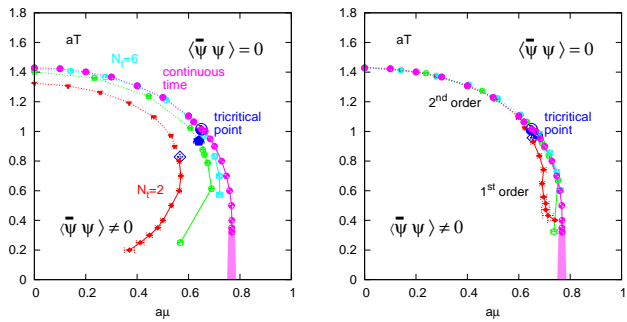}\\
\caption{
{\em Left:} Lattice QCD phase diagram in the strong coupling limit, setting $a/a_t=\gamma^2$ following mean-field.
Different results are obtained for different numbers $N_t$ of time-slices: $N_t=2$, $N_t=4$~\cite{Fromm2010}, $N_t=6$
and $N_t=\infty$ (i.e. continuous Euclidean time)~\cite{Unger2011}. {\em Right:} Same, with corrected anisotropy
$a/a_t = \gamma^2 \exp(c/\gamma^2)$ and ${\cal O}(1/N_t)$ corrections. All results coincide. The re-entrance
at low $aT$ is a finite-$N_t$ artifact.
}
\label{PhaseDiagSC}
\vspace{-4mm}
\end{figure}

This system has been studied since decades, both via mean field \cite{Kawamoto1981,Damgaard1985,Bilic1992a,Bilic1992b,Nishida2004,Ohnishi2009} 
and by Monte Carlo methods~\cite{Wolff1985,Karsch1989,Fromm2010}. 
In recent years, the latter 
have undergone a revival using 
the Worm algorithm \cite{Adams2003,Fromm2010,Unger2011}, 
which violates the Grassmann constraint in order to sample the monomer two-point function $G(x,y)$, from which 
the chiral susceptibility can be obtained. 
These techniques have been applied to obtain all lattice data presented here. 
We study the chiral limit $m_q=0$ which does not incur a penalty in computer cost, contrary to the usual determinantal approach. The staggered action $S_F$ Eq.~(\ref{QCDAction})  then 
satisfies a $U(1)$ ``remnant'' chiral symmetry, which is spontaneously broken at low temperature and density,
with order parameter $\expval{\pbp}$.
In Fig.~\ref{PhaseDiagSC} left, we show the ($\mu,T$) phase diagram in the strong-coupling (SC) limit.
It is qualitatively similar to the expected phase diagram of QCD in the chiral limit: the transition is of 
second order from $a\mu=0$ up to a tricritical point $(a\mu_T,aT_T)$, then turns first order. 
At finite quark mass, the second order line turns into a crossover and the tricritical point into a second order 
critical endpoint. 
Note the different phase boundaries obtained from lattices with different numbers $N_t$ of time-slices:
they converge to the continuous-time phase boundary as $N_t\to\infty$. But the $N_t$-dependence is strong,
and there are indications of re-entrance at low temperature, supported by mean-field calculations~\cite{Nishida2004},
which only disappear in the continuous-time limit. There are two reasons for this: 
$(i)$ the transition temperature is subject to 
${\cal O}(1/N_t)$ corrections, as studied in \cite{Unger2011};
$(ii)$ temperatures $aT > 1/2$ can only be explored by using anisotropic lattices where $a/a_t > 1$. 
But the relationship
between the bare anisotropy $\gamma$ in the action $S_F$ Eq.(\ref{QCDAction}) and $a/a_t$ is not known exactly. Mean-field indicates $a/a_t \!=\! \gamma^2$, which was used
to obtain Fig.~\ref{PhaseDiagSC} left. Assuming the form  $a/a_t \!=\! \gamma^2 \exp(c / \gamma^2)$ and allowing for small
${\cal O}(1/N_t)$ corrections (Fig.~\ref{PhaseDiagSC} right) produces much more consistent results.

A crucial question is whether this phase diagram develops qualitatively new features as $\beta$
is increased from $0$ to $\infty$. At low temperature especially, things may change: when $\beta\!=\!0$,
the transition at $\mu_c(T\!=\!0)$ separates a chirally-broken, baryon-free vacuum and a chirally-symmetric, 
baryon-saturated state with one static baryon per lattice site. That is a very crude cartoon of a nuclear
matter phase: in the continuum limit, depending on the chemical potential, it may evolve into a nuclear
liquid, a crystalline phase, a color superconductor, etc... A first insight may be gained by considering
${\cal O}(\beta)$ corrections to the $\beta\!=\!0$ phase diagram. At the same time, we can also address an interesting 
quantitative issue: the ratio $T_c(\mu\!=\!0)/\mu_c(T\!=\!0)$ is about (160 MeV)/(300 MeV) $\sim 0.53$ in nature, but
about $1.402/0.75 \approx 1.87$ when $\beta\!=\!0$. How does it vary with $\beta$ ?


\smallskip

\emph{Corrections to the Strong Coupling Limit -}
To go beyond the strong coupling limit, a systematic expansion of the QCD partition 
function in $\beta$ is needed, which we perform to first order $\Ord{\beta}$. 
Writing the $\beta\!=\!0$ partition function as $Z_{SC}\!=\!\int \dpdpb Z_F$, with
$Z_{F}(\psi,\bar{\psi})\!=\!\int dU e^{S_F}$ the fermionic partition function, the $\beta\!\neq\! 0$ partition function Eq.~\ref{QCDPF} becomes:
\begin{eqnarray}
\hspace*{-0.3cm}
Z_{\rm QCD}=\int \dpdpb dUe^{S_F+S_G}=\int \dpdpb Z_{F} \expval{e^{S_G}}_{Z_F} ,\\
\expval{e^{S_G}}_{Z_F}\simeq 1\!+\! \expval{S_G}_{Z_F}=1 \!+\! \frac{\beta}{2N_c} \sum_P \expval{\tr[U_P+U_P^\dagger]}_{Z_F}, \label{eq:cumulant}
\label{ObetaPF}
\end{eqnarray}

\noindent
where Eq.~(\ref{ObetaPF}) is an $\Ord{\beta}$ truncation.
We thus need
the expectation value of the elementary plaquette $\tr[U_P]$ in the strong coupling ensemble $Z_F$. 
The plaquette is composed of 4 links representing gluons,
which provide new possibilities to make
color singlets together with $\bar{\psi}_x \psi_{x\pm\hat{\mu}}$ propagating fermions.

This gives rise (for $N_c\!=\!3$) to 19 terms, which are computed
from the product $P=J_{ij}J_{jk}J_{kl}J_{li}$ of the one-link integrals 
$J_{ij} \equiv \int dU U_{ij} \exp(\bar\psi U \phi - \bar\phi U^\dagger \psi)$
around an elementary plaquette \cite{Creutz1978,Azakov1988,Jens2009} :
\begin{eqnarray}
J_{ij}&=&-\sum_{k=1}^{3}\frac{(3-k)!}{3!(k-1)!}
\Big[M_\psi M_\phi\Big]^{k-1}\bar{\phi}_j\psi_i  \nonumber \\
&+&\frac{1}{12}\varepsilon_{ii_2i_3}
\varepsilon_{jj_2j_3} \bar{\psi}_{i_2}\phi_{j_2}\bar{\psi}_{i_3}\phi_{j_3}
- \frac{1}{3}\bar{B}_\psi B_\phi \bar{\phi}_j\psi_i\;,
\label{LinkIntegral}
\end{eqnarray}

\noindent
where $M$ and $B$ represent mesons and baryons.
The first term describes the propagation of a $(\bar{q} g)$ anti-quark+gluon together
with $0$ to $2$ mesons; the second term describes a $(qqg)$; the third term is
again a $(\bar{q} g)$ together with a baryon.
From these, we compute the weight associated with a plaquette (or any Wilson loop) 
source term in the strong coupling configuration. 

At the corners of the plaquette, the Grassmann variables $\psi,\phi$ 
are bound into baryons and mesons to fulfill the Grassmann constraint Eq.(\ref{Grassmann}), giving
rise to the 19 subgraphs mentioned above.
Introducing a variable $q_P \in \{0,1\}$ to mark the ''excited'' plaquettes $P$ associated with
the second term of Eq.(\ref{eq:cumulant}), and corresponding variables $q_b$ and $q_x = q_P$
for the links and the corners of such plaquettes, we can write the ${\cal O}(\beta)$ partition
function in the same form as Eq.(\ref{eq:Z_SC}) with modified weights $\hat{w}$:
\begin{align}
Z(\beta) &=\sum_{\{n,k,\ell,q_P\}}\prod_x \hat{w}_x \prod_{b} \hat{w}_b \prod_\ell \hat{w}_\ell \prod_P \hat{w}_P \\
\hat{w}_x & = w_x v_x,\qquad \hat{w}_b = w_b k_b^{q_b}, \\
\hat{w}_\ell& = w_\ell \prod_\ell { w_{B_i}(\ell)}, \qquad
\hat{w}_P=\left(\frac{\beta}{2N_c}\right)^{q_P}\;,
\end{align}
where $v_x\!=\!(N_c-1)!$ if $x$ is the corner of an excited plaquette attached to an external
meson line, $N_c!$ if it is attached  to an external baryon line, $1$ otherwise.
Likewise, the weight of each baryon loop segment $l$ is modified by a factor $w_{B_1}=\frac{1}{({\Nc-1)!}}$, $w_{B_2}=(\Nc-1)!,$
where $B_1$ and $B_2$  correspond to the second/third expression in Eq.(\ref{LinkIntegral}).
We can sample this partition function by the same worm algorithm as for $\beta\!=\!0$, adding a Metropolis
step to update the plaquette variables $q_P$.
In practice, we found it simpler to obtain gauge observables via reweighting from the $\beta\!=\!0$ ensemble.

Several qualitatively new features are made possible by including $\Ord{\beta}$ contributions:
$(i)$ the constituent quarks of baryons and mesons can now separate: hadrons are no longer point-like, 
but acquire a size $\sim a$;
$(ii)$ the baryon-baryon interaction can now proceed by quark exchange: it is no longer limited to 
the on-site Pauli exclusion principle;
$(iii)$ baryon saturation can now coexist with monomers, making chiral symmetry breaking possible in 
the dense phase similar to nuclear matter.

\begin{figure}
\includegraphics[width=0.48\textwidth]{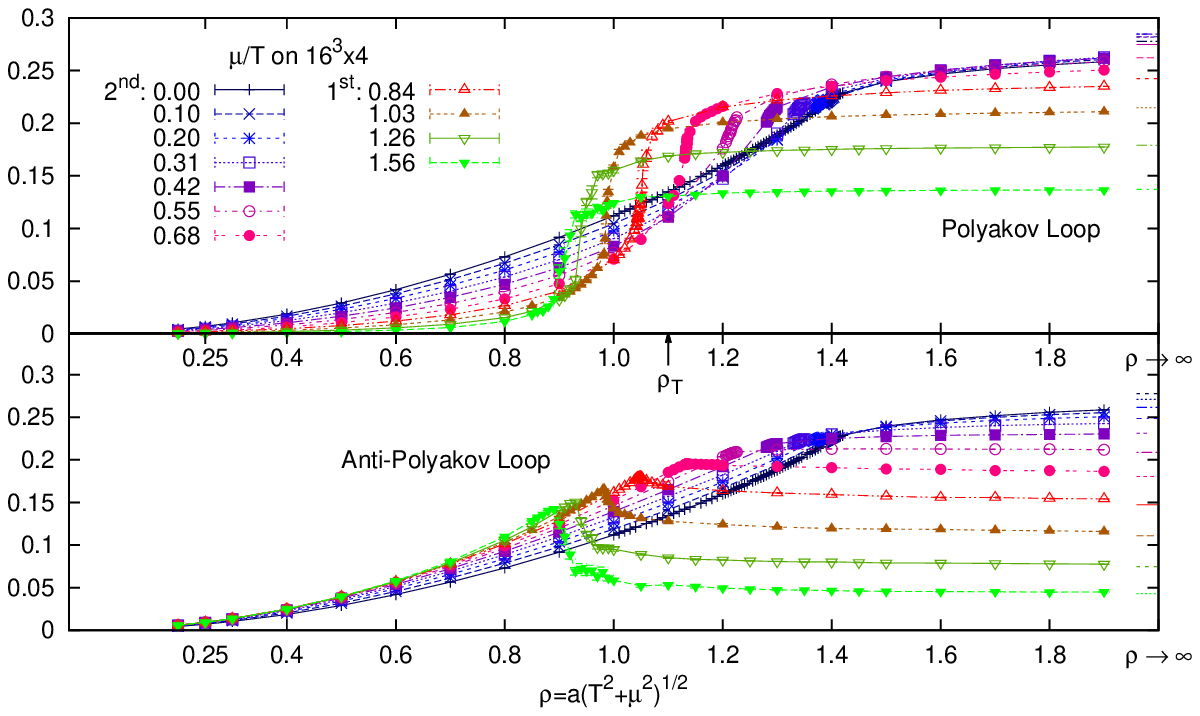}
\caption{Polyakov loop $\frac{1}{3}\expval{\tr L}$ and anti-Polyakov loop $\frac{1}{3}\expval{\tr L^*}$ 
as a function of $(\mu,T)$ on a $16^3\times 4$
lattice at $\beta=0$. The colors label successive values of $\mu/T$, and the $x$-axis is
$\rho \equiv a \sqrt{\mu^2+T^2}$. At the tricritical point, $\rho_T =  1.10(2).$
}
\label{mudepPol}
\vspace{2mm}
\includegraphics[width=0.48\textwidth]{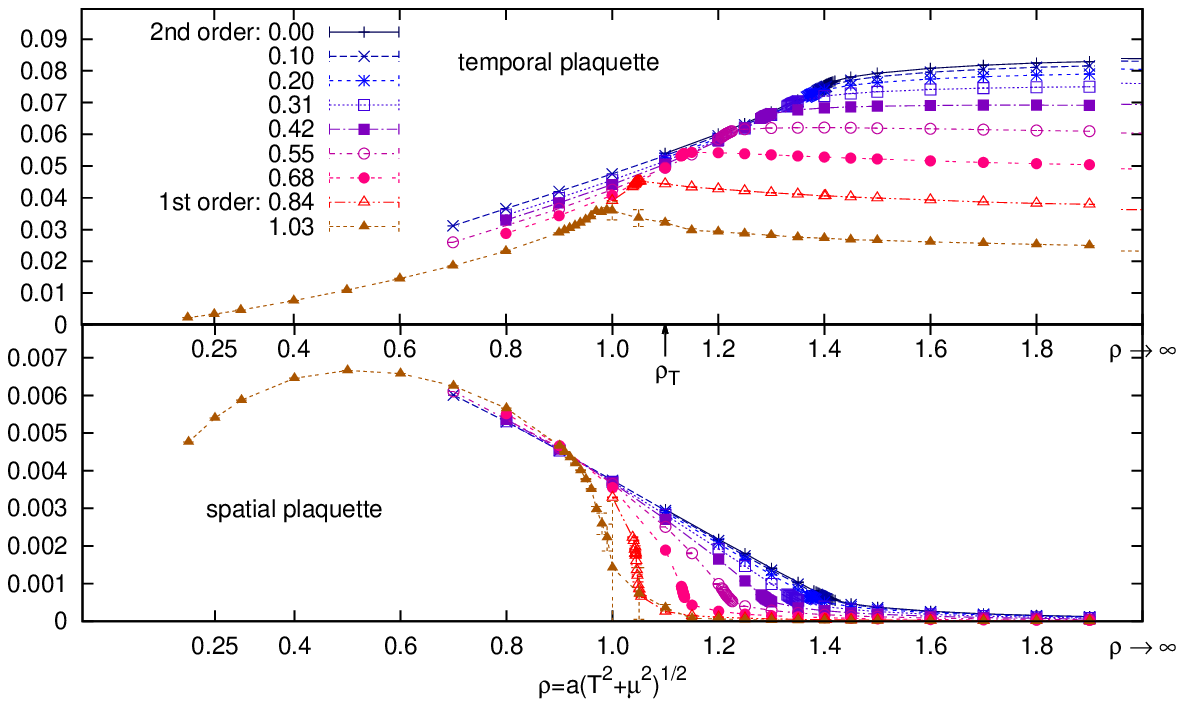}
\vspace{-5mm}
\caption{Average spatial/temporal plaquette 
$\frac{1}{3}\expval{\tr P_s}$, $\frac{1}{3}\expval{\tr P_t}$ as a function of $(\mu,T)$ on a $16^3\times 4$
lattice at $\beta=0$. 
Wilson loops are sensitive to the chiral transition 
and develop a discontinuity as the transition turns first order.
$\expval{\tr P_s}$ varies oppositely to $\expval{\tr P_t}$, and remains very small.
}
\label{mudepPlaq}
\vspace{-5mm}
\end{figure}

\smallskip

\emph{Wilson loops at $\beta\!=\!0$ -}
Figs.~\ref{mudepPol} and \ref{mudepPlaq} illustrate the dependence of the Polyakov loop and of the plaquette (time-like
and space-like) on the chemical potential $\mu$ and the temperature $T$, at $\beta\!=\!0$.
The $x$-axis represents the ``distance'' $a\sqrt{\mu^2+T^2}$ from the vacuum, and 
different symbols are used for different values of $\mu/T$.
Several features are noticeable: $(i)$ the plaquette has a non-zero value, even at $T\!=\!\mu\!=\!0$.
This is caused by the ordering effect of the fermions. Indeed, increasing the number of quark
fields from 1 to 13 triggers restoration of the chiral symmetry~\cite{Forcrand2013}.
$(ii)$ The first-order phase transition is visible at large $\mu/T$ through a discontinuity in all
Wilson loops, although it is associated with breaking/restoration of the chiral symmetry.
This can be assigned to the non-zero latent heat.
$(iii)$ Even in the regime of small $\mu/T$, where the chiral transition is second-order,
the Polyakov loop is clearly sensitive to the transition as already found in U(3) gauge theory~\cite{Fromm2011}, and reflecting the
''entanglement'' of confinement and chiral symmetry breaking seen in effective models~\cite{Weise}.

\smallskip

\emph{Phase Diagram as a Function of $\beta$ -}
We now show how to obtain the derivative $d(aT_c)/d\beta$ of the chiral transition temperature $aT_c$ with respect to $\beta$. 
Since the worm algorithm samples the 2-point correlation function $G(x_1,x_2)$, we can measure its integral, 
which is equal to the chiral susceptibility $\chi$ (there is no disconnected piece $\expval{\pbp}^2$: 
since we set $m_q=0$ and work in a finite volume,
$\expval{\pbp}=0$ both in the chirally symmetric and broken phase),
\begin{align}
 \chi \equiv \expval{(\pbp)^2} = \frac{1}{L^3 N_\tau}\sum_{x_1,x_2} G(x_1,x_2) .
\end{align}

\begin{figure}[t!]
\centerline{\includegraphics[width=0.48\textwidth]{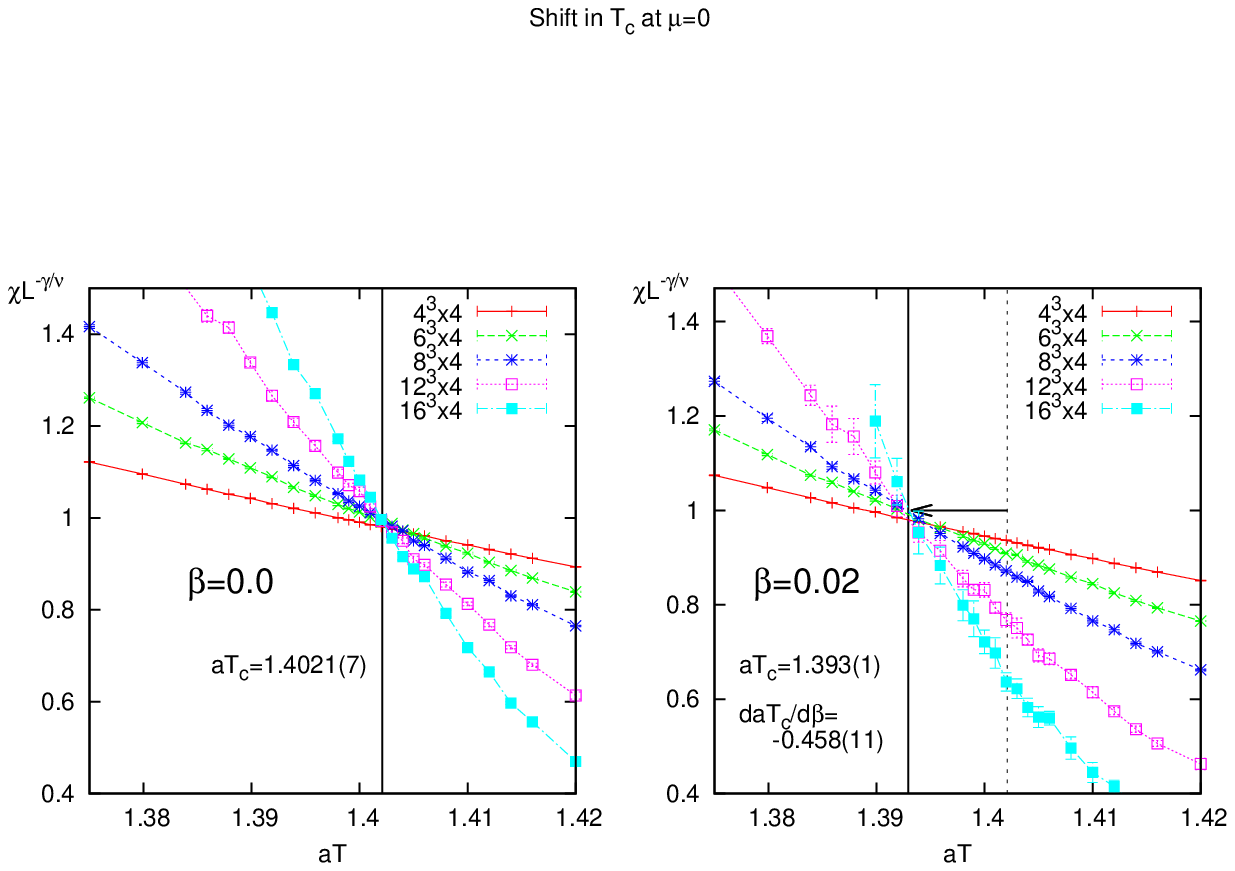}}
\caption{
The $\mu\!=\!0$ transition temperature $aT_c$ from finite-size scaling of the chiral susceptibility on $N_t=4$ lattices. {\em Left:} $\beta=0$. {\em Right:} $\beta=0.02$.
The arrow marks the shift in $aT_c$.}
\label{susc}
\end{figure}

\noindent
At $\beta\!=\!0$ and for some $\mu < \mu_T$, the critical temperature $aT_c(\mu)$ can be obtained
from finite-size scaling: the curves $\chi(aT,L) L^{-\gamma/\nu}$ obtained on several lattice sizes $L$ all intersect
at $T\!=\!T_c(\mu)$, with a slope $\propto L^{1/\nu}$ at the intersection, as illustrated Fig.~\ref{susc} left.
The transition is in the $3d~O(2)$ universality class with known critical exponents, which facilitates
the analysis. In the region of a first-order transition, $\mu > \mu_T$, this ansatz is modified accordingly,
following Ref.~\cite{Borgs-Kotecky}.
When we turn on $\beta$, the chiral susceptibility changes, and we can easily measure its derivative, since 
\begin{align}
\frac{d\chi}{d\beta} = 3 L^3\Nt\expval{(\pbp)^2 P_t }-\expval{(\pbp)^2}\expval{P_t}.
\label{eq:dchi_dbeta}
\end{align}
While both the temporal and the spatial plaquettes formally enter in this expression, the latter 
is a factor $\gtrsim 10$ smaller than the former, cf.~Fig.~\ref{mudepPlaq}.
The effect of $\beta$, to linear order, is illustrated Fig.~\ref{susc} right. At temperature $aT_c$,
the rescaled chiral susceptibility $\chi L^{-\gamma/\nu}$ changes by $\beta \frac{d\chi}{d\beta} L^{-\gamma/\nu}$.
This $L$-dependent change produces a horizontal shift of the intersection point
\begin{align}
\beta \frac{d(aT_c)}{d\beta} = \beta \frac{d\chi}{d\beta} L^{-\gamma/\nu} [d\chi/d(aT)]^{-1}.
\end{align}
The $L$-independence of this shift
is a consistency check of our analysis.
The highest accuracy is achieved when $\mu\!=\!0$, for which we determine (on $N_t\!=\!4$ lattices):\linebreak
$\left.aT_c\right|_{\beta=0} = 1.4021(7)$,$\left.\frac{d}{d\beta}aT_c(\beta)\right|_{\beta=0} \!\!= -0.46(1)$.

\begin{figure}[t!]
\vspace{-2mm}
\centerline{
\includegraphics[width=0.48\textwidth]{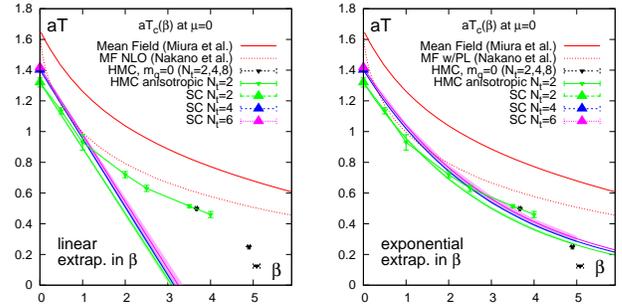}
}
\caption{
Phase boundary in $\beta-aT$ plane at $\mu=0$. \emph{Left:} linear extrapolation. \emph{Right:} exponential extrapolation. 
The boundary coincides very well with conventional Hybrid Monte Carlo data at large $\beta$. Also, the phase boundary is 
rather similar to the one obtained via
a mean field theory prediction without \cite{Miura2009} and with Polyakov loop effects \cite{Nakano2010}.
\vspace{-2mm}
}
\label{Miura}
\end{figure}

Several observations are in order.
First, $aT_c$ decreases as $\beta$ increases: this is as it should be, since $a$ decreases.
Secondly, this $\Ord{\beta}$ result can be compared with mean-field predictions \cite{Miura2009,Nakano2010}, 
see Fig.~\ref{Miura}. The agreement is rather good.
More importantly, at $\mu\!=\!0$ we can compare with finite-$\beta$ Hybrid Monte Carlo simulations (at $\mu\!=\!0$, these simulations
are sign-problem free) performed on $N_t\!=\!2$
and $N_t\!=\!4$~\cite{Gottlieb1987,Gavai1990,DElia2003} lattices
with isotropic actions (i.e. $aT \!=\! 1/2$ and $1/4$, respectively) and extrapolated to zero quark mass. 
These data points are marked in black in Fig.~\ref{Miura}. 
We have also computed $aT_c(\mu=0)$ ourselves, using HMC on anisotropic lattices. 
As Fig.~\ref{Miura} left shows, our $\Ord{\beta}$ determination of $aT_c(\mu=0)$ agrees perfectly with the linear approximation to the HMC determination. But the latter shows significant curvature. To better approximate the exact result, we perform an empirical, exponential extrapolation
$aT_c(\mu=0,\beta) / aT_c(\mu=0,\beta=0) \approx \exp (\beta \frac{d}{d\beta} aT_c|_{\beta=0})$.
As seen in Fig.~5 right, it turns out that this approximation,
which includes a resummation of higher-order $\beta$-contributions,
follows the exact HMC result up to $\beta\sim 5$ (or $a\sim 0.3$ fm),
where the lattice theory is much closer to continuum physics. 
We have applied the same procedure to determine $aT_c(\beta)$ at non-zero chemical potential.
Although the statistical errors increase and the scaling window shrinks,
$d(aT_c)/d\beta$ is clearly not as large as when $\mu\!=\!0$. In fact, $d(aT_c)/d\beta$ becomes consistent with zero as $\mu$ approaches $\mu_T$. The tricritical point and the first order line
seem to only weakly
depend on $\beta$. Thus, $T_c(\mu=0)/\mu_c(T=0)$ decreases at $\mathcal{O}(\beta)$ towards its continuum value.

\begin{figure}
\centerline{
\includegraphics[width=0.48\textwidth]{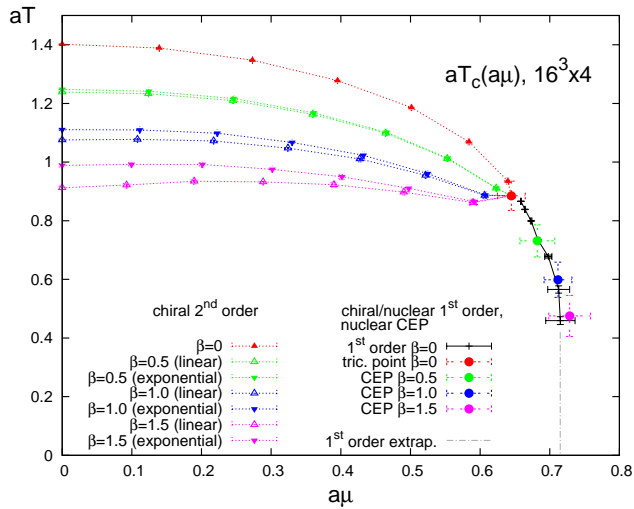}
}
\caption{
Phase boundary in the $\mu-T$ plane in the strong coupling limit and extrapolated to finite $\beta$, comparing linear and exponential extrapolation. 
We do not observe a shift of the chiral tricritical point.
The nuclear critical endpoint (CEP), determined from 
the reweighted baryon density, moves down along the
first order line (extrapolated to $T=0$ to guide the eye) as $\beta$ is increased.
}
\label{finiteMu}
\end{figure}
\begin{figure}
\centerline{
\includegraphics[width=0.45\textwidth]{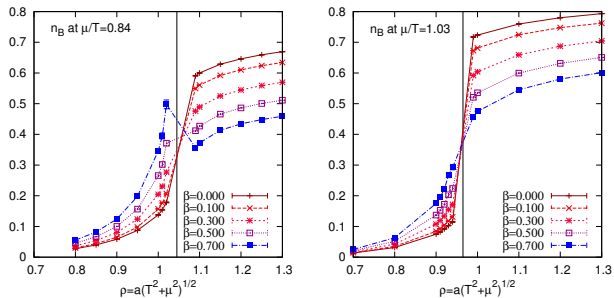}}
\caption{Reweighted baryon density $n_B$ for $\mu/T > \mu_T/T_T \approx 0.71$, i.e.~in the
first-order regime. The nuclear transition weakens as $\beta$ is increased. 
At some $\beta_c$ it turns from first order to second order, when the jump in the
baryon density vanishes.
The larger $\mu/T$, the stronger the first order transition, and the larger $\beta_c$. \emph{Left:} $\mu/T$ is close to the tricritical point, $\beta_c\approx 0.3$. 
\emph{Right:} $\mu/T$ is larger and $\beta_c\approx 0.7$.
}
\label{baryondensity}
\end{figure}

The resulting phase diagram is illustrated in Fig.~\ref{finiteMu} for $\beta\!=\!0.5$, $1.0$ and $1.5$.
We show the phase boundary obtained by linear reweighting, based on Eq.~(\ref{ObetaPF}), compared to the one obtained by exponential extrapolation,
which works so well at $\mu=0$. In both cases, the phase boundary becomes more ``rectangular'' as one moves away from the strong
coupling limit: the second-order transition line becomes ``flatter'' (less $\mu$-dependent),
and the first-order transition line remains almost ``vertical'', leaving the tricritical point
at the ``corner of the rectangle''. From the chiral susceptibility, no clear shift of $(a\mu_T,aT_T)=(0.65(2),0.91(5))$ could be detected; however, from the 
baryon density $n_B$, see Fig. \ref{baryondensity}, we have evidence 
that the critical endpoint of the nuclear transition, which coincides with the chiral transition at $\beta=0$, moves along the first order line, to smaller values of $T$. 
This is expected: as $\beta$ increases, the lattice spacing $a$ shrinks,
and $(a M_B)$ also, where $M_B$ is the baryon mass. If $(a \mu_c)$ stays approximately constant
as we observe, then the nuclear attraction responsible for the 
difference $(M_B - 3 \mu_c(T=0))$, of about 300 MeV when $\beta=0$~\cite{Fromm2010},
becomes weaker. The weakening of the associated first-order transition
brings the nuclear critical endpoint point down in temperature.
We plan to study $\Ord{\beta^2}$ corrections next.

We would like to thank K.~Miura and A.~Ohnishi for helpful discussions.


\begin{thebibliography}{99}

\bibitem{Forcrand2009}
  P.~de Forcrand,
  PoS LAT {\bf 2009} (2009) 010

  \bibitem{Fromm2011b}
  M.~Fromm, J.~Langelage, S.~Lottini and O.~Philipsen,
  JHEP {\bf 1201} (2012) 042
  
  \bibitem{Fromm2012}
  M.~Fromm, J.~Langelage, S.~Lottini, M.~Neuman and O.~Philipsen,
  Phys.\ Rev.\ Lett.\  {\bf 110} (2013) 122001

  \bibitem{Rossi1984}
P.~Rossi, U.~Wolff,
{\em Nucl. Phys. B {\bf 258}} (1984) 105.

  \bibitem{Wolff1985}
U.~Wolff,
{\em Phys. Lett. B {\bf 153}} (1985) 92.
  
\bibitem{Fromm2010}
  P.~de Forcrand and M.~Fromm,
  Phys.\ Rev.\ Lett.\  {\bf 104} (2010) 112005
  [arXiv:0907.1915 [hep-lat]].
  

\bibitem{Mercado2013}
  Y.~D.~Mercado, C.~Gattringer and A.~Schmidt,
  Phys.\ Rev.\ Lett.\  {\bf 111} (2013) 141601
  [arXiv:1307.6120 [hep-lat]].

\bibitem{Eriksson1981}
K.~E.~Eriksson, N.~Svartholm, B.~S.~Skagerstam
{\em J. Math. Phys. {\bf 22}} (1981) 2276.


\bibitem{Kawamoto1981}
  N.~Kawamoto and J.~Smit,
  Nucl.\ Phys.\ B {\bf 192} (1981) 100.

  \bibitem{Damgaard1985}
  P.~H.~Damgaard, D.~Hochberg and N.~Kawamoto,
  Phys.\ Lett.\ B {\bf 158} (1985) 239.

  \bibitem{Bilic1992a}
  N.~Bilic, F.~Karsch, K.~Redlich,
  {\em Phys. Rev. D {\bf 45}} (1992) 3228.


\bibitem{Bilic1992b}
  N.~Bilic, K.~Demeterfi, B.~Petersson,
  {\em Nucl. Phys. B {\bf 377}} (1992) 3651.


\bibitem{Nishida2004}
Y.~Nishida,
{\em Phys. Rev. D {\bf 69}} (2004) 094501.


\bibitem{Ohnishi2009}
A.~Ohnishi, K.~Miura, T.~Nakano, N.~Kawamoto,
{\em PoS (LAT2009)} 160.

\bibitem{Karsch1989}
F.~Karsch, K.~H.~M\"utter,
{\em Nucl. Phys. B {\bf 313}} (1989) 541.
  

\bibitem{Adams2003}
  D.~H.~Adams and S.~Chandrasekharan,
  {\em Nucl.\ Phys.\  B {\bf 662}} (2003) 220.
  
  
\bibitem{Unger2011}
  W.~Unger and P.~de Forcrand,
  PoS LATTICE {\bf 2011} (2011) 218

\bibitem{Azakov1988}
  S.~I.~Azakov and E.~S.~Aliev,
  Phys.\ Scripta {\bf 38} (1988) 769.

\bibitem{Creutz1978}
  M.~Creutz,
  J.\ Math.\ Phys.\  {\bf 19} (1978) 2043.

  \bibitem{Jens2009}
  Jens Langelage, PhD thesis (2009)

\bibitem{Forcrand2013}
  P.~de Forcrand, S.~Kim and W.~Unger,
  JHEP {\bf 1302} (2013) 051
  [arXiv:1208.2148 [hep-lat]].

  
\bibitem{Fromm2011}
  M.~Fromm, J.~Langelage, O.~Philipsen, P.~de Forcrand, W.~Unger and K.~Miura,
  PoS LATTICE {\bf 2011} (2011) 212
  [arXiv:1111.4677 [hep-lat]].
  
\bibitem{Weise}
  T.~Hell, K.~Kashiwa and W.~Weise,
  Phys.\ Rev.\ D {\bf 83} (2011) 114008
  [arXiv:1104.0572 [hep-ph]].
  
\bibitem{Borgs-Kotecky}
C.~Borgs and R.~Kotecky, 
J.\ Stat.\ Phys.\ {\bf 61} (1990) 79.
   
\bibitem{Miura2009}
  K.~Miura, T.~Z.~Nakano, A.~Ohnishi and N.~Kawamoto,
  Phys.\ Rev.\ D {\bf 80} (2009) 074034

\bibitem{Nakano2010}
  T.~Z.~Nakano, K.~Miura and A.~Ohnishi,
  Phys.\ Rev.\ D {\bf 83} (2011) 016014
  [arXiv:1009.1518 [hep-lat]].
  
  \bibitem{Gottlieb1987}
  S.~A.~Gottlieb, W.~Liu, D.~Toussaint, R.~L.~Renken and R.~L.~Sugar,
  Phys.\ Rev.\ D {\bf 35} (1987) 3972.

  \bibitem{Gavai1990}
  R.~V.~Gavai {\it et al.}  [MT(c) Collaboration],
  Phys.\ Lett.\ B {\bf 241} (1990) 567.


 \bibitem{DElia2003}
  M.~D'Elia and M.~-P.~Lombardo,
  Phys.\ Rev.\ D {\bf 67} (2003) 014505
  [hep-lat/0209146].

    
\end{thebibliography}
\end{document}